\documentclass[twocolumn,showpacs,preprintnumbers,amsmath,amssymb]{revtex4-1}

\usepackage{graphicx}% Include figure files
\usepackage{dcolumn}% Align table columns on decimal point
\usepackage{bm}% bold math

\begin{document}

\title{Geometrical Aspects of Venus Transit \hspace{15cm}\small{(Aspectos Geom\'etricos do Tr\^ansito de V\^enus)}}
\author{A. C. Bertuola$^{1}$, C. Frajuca$^{1}$, N. S. Magalh\~{a}es$^{2}$, V. S. Filho$^{3}$}

\affiliation{$^{1}$S\~ao Paulo Federal Institute, S\~ao Paulo, 01109-010, Brazil}
\affiliation{$^{2}$S\~ao Paulo Federal University, S\~ao Paulo, 04021-001, Brazil}
\affiliation{$^{3}$H4D Scientific Research Laboratory, 04674-225, S\~ao Paulo, SP, Brazil}

\begin{abstract}

\begin{center}

{\bf{Abstract}}

\end{center}

We obtained two astronomical values, the Earth-Venus distance and Venus
diameter, by means of a geometrical treatment of photos taken of Venus
transit in June of 2012. Here we presented the static and translational models
that were elaborated taking into account the Earth and Venus orbital 
movements. An additional correction was also added by considering the Earth
rotation movement. The results obtained were compared with the 
values of reference from literature, showing very good concordance.\\

{\bf{Keywords:}} Venus transit, geometrical methods, Earth-Venus distance, Venus diameter

\begin{center}

{\bf{Resumo}}

\end{center}

N\'os obtivemos dois valores astron\^omicos, a dist\^ancia Terra-V\^enus e o di\^ametro de V\^enus, 
por meio de um tratamento geom\'etrico de fotos tiradas do tr\^ansito de V\^enus em junho de 2012. 
Aqui n\'os apresentamos os modelos est\'atico e translacional que foram elaborados levando-se em conta 
os movimentos orbitais da Terra e de V\^enus. Uma corre\c c\~ao adicional foi tamb\'em acrescentada, 
considerando o movimento de rota\c c\~ao da Terra. Os resultados obtidos foram comparados com os valores 
de refer\^encia da literatura, mostrando uma concord\^ancia muito boa. \\

{\bf{Palavras-chave:}} Tr\^ansito de V\^enus, m\'etodos geom\'etricos, dist\^ancia Terra-V\^enus, di\^ametro de V\^enus

 \pacs{96.30.Ea, 95.10.Eg, 96.15.De}
% 96.30.Ea Venus
% Orbits determination of, 95.10.Eg
% Planets orbits and rotation, 96.15.De
\end{abstract}

\maketitle

\section{Introduction}
\label{intro}
In June of 2012, in some places of the Earth some people had the privilege to witness Venus 
transit~\cite{Dick2004,Pasachoff2012}, an astronomical event of singular beauty, observed and studied for a 
long time~\cite{Hind1871,Forbes1879,Argyll+1882,Meadows1974}. This event is very rare and it is possible that this generation of researchers does not again see the phenomenon because the next one will
occur only in 2117. The phenomenum is very important in order to obtain some physical informations of the planet. For instance, spectrographic data can be combined with refraction measurements from Venus transits to give a scale height of the scattering centers in Venus haze~\cite{Goody1967}. Further in a recent work.~\cite{Chassefierea2012}, it is proposed that accurate astronomical distances may be determined from recent observations which were collected in the last event. 

In the literature, investigations concerning to calculations of orbits and astronomical observables associated to them have been considered in many contexts with very sophisticated theoretical frameworks~\cite{Heilmann,Suspes,Bolos,Pugliese}, including aspects like either the sensibility to disturbances in the orbit of satellites~\cite{Heilmann} or  
the study of equatorial circular orbits in static axially symmetric gravitating systems~\cite{Suspes} and the calculation 
of radius of the orbits as well. In general, relativistic frameworks have been used~\cite{Bolos,Pugliese}. Here we intend 
to show that it is possible in a very simple case to calculate some astronomical observables with a simple geometrical 
method and elementary algebra, by solely considering the photos of Venus transit.  

A lot of photos were taken of the phenomenon and published by several different authors~\cite{Mosaic2012}. The following illustration displays a mosaic composed by a selection of photos that were chosen for our study.
%%%%%%%%%%%%%%%%%%%%%%%% FIGURE 1 %%%%%%%%%%%%%%%%%%%%%%%%%%%%%%%%%%%%%%
 \begin{figure}[ht]
\begin{centering}
 \includegraphics[width=7.5cm]{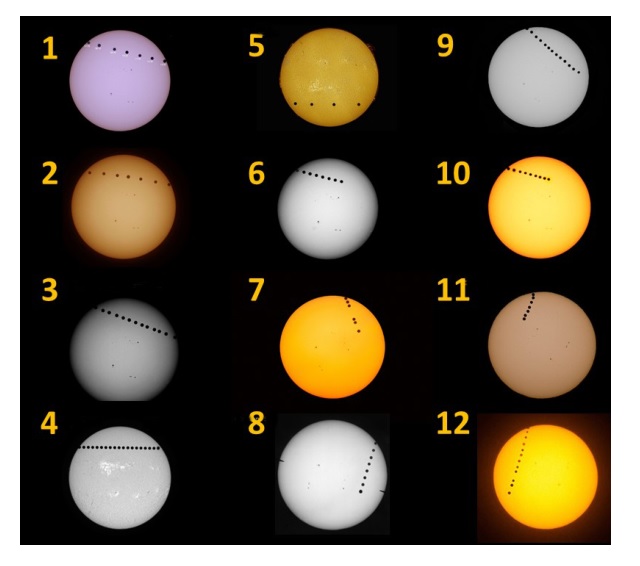}
 \caption[dummy0] {Mosaic formed by photos~\cite{Mosaic2012} with the 
 composition of the positions of Venus during the transit. From frames 1 up to 5, one 
 observes the complete composition of the transit. For the others remaining 
 photos, the composition of the transit is incomplete.} 
\end{centering}
\end{figure}
%%%%%%%%%%%%%%%%%%%%%%%%%%%%%%%%%%%%%%%%%%%%%%%%%%%%%%%%%%%%%%
The Fig.~1 is a mosaic constructed from  several photos~\cite{Mosaic2012} obtained in different
places. In the frames 1 up to 5, Venus transit can be seen in a complete
way. The other photos (6 up to 12) are those that have the incomplete
composition because the unset happened before the end of the transit.
Working directly with the photos of Fig. 1, there is the possibility to
obtain two ratios: the length of the trajectory of the spot by the Sun 
diameter and the diameter of Venus spot by the Sun diameter. 

Our work was initially inspired in Ref.~\cite{Stone+1882}, in which different methods could be constructed to accurately calculate the Sun's distance from a discussion of observations of Venus transit, so that British expeditions could be organized. We follow an inverse way, that is, known the Earth-Sun distance, we obtained the Earth-Venus distance and Venus diameter analyzing the transit phenomenon. 
 
We constructed two physical models in order to calculate Venus diameter and the Earth-Venus distance. 
The first model considered is the static case, in which the Earth and the Sun are in rest; the second one is the translation model, in which the translational movement of the Earth is taken into account, but it is not considered the known spin of the planets~\cite{Hughes2003}. Additionally, the latter had a correction due to the rotation of the Earth. We used in our calculations some physical quantities known~\cite{Halliday2012,pdg2010,Brown1998,Espenak2014}, as shown in the table 1. In that table, $d_{ES}$ is the maximum distance from the Earth to the Sun (The Earth is close to the apogee.) and $D_{S}$ is the Sun  diameter. Here we consider the diameter as the Sun equatorial diameter~\cite{Brown1998}. The other astronomical values are the Earth orbital period $(T_{E})$, Venus orbital period $(T_{V})$, the value of the time spent for Venus to complete the 
transit $(t)$ and the radius of the Earth $(R_{E})$.  The official values of the Earth-Venus distance and Venus diameter used for 
comparison in this work are respectively $d_{EV}=4.31 \times {\small 10}^{7}$ km and $D_{V}=1.2104 \times {\small 10}^{4}$~km~\cite{Halliday2012,pdg2010,Brown1998,Espenak2014}. 
\begin{table}[h]
\begin{center}
\caption{The quantity $d_{ES}$ is the Earth-Sun distance (in km), 
$D_{S}$ is the Sun diameter, $T_{E}$ is the Earth revolution  
period (in days), $T_{V}$ is Venus orbital period (in days), 
$t$ is the duration of transit (in hours) and $R_{E}$ is the radius of the Earth (in km).}
\begin{tabular}{|c|c|c|c|c|c|}
\hline
$d_{ES}(km)$ & $\frac{d_{ES}}{D_{S}}$ & $T_{E}(day)$ & $T_{V}(day)$ & $t\left(h\right)$ & $R_{E}$$\left(km\right)$ \\ 
\hline
$1.521$x$10^{8}$ & $109.3$ & $365.26$ & $224.7$ & $6.67$ & $6378$ \\ 
\hline
\end{tabular}
\end{center}
\label{table1}
\end{table}

The numerical values of the Earth-Venus distance and Venus diameter were
calculated when Venus was also farthest away from the Sun. In the following, 
we describe the geometrical methods that we have elaborated. 

\section{Geometrical Aspects}

The direct geometric method is based on a technique of image treatment of photos of
Venus transit. By means of this technique, two important ratios are
obtained. The first ratio is the length ($L^\prime $) of the trajectory of
Venus spot on the Sun by the Sun diameter. The second ratio is the
value of the diameter of Venus spot ($D^\prime $) by the Sun diameter. The
indirect geometric method corresponds to a mathematical triangulation
technique of the relative positions of the Earth, the Sun and Venus during
the transit phenomenon.  The Venus diameter and its distance to the Earth were 
obtained by combining those two methods. In the following, we describe
details of both geometrical methods.
\subsection{Direct Geometrical Method}
We used the Photoshop software, so that the coordinates x and y of any point
can be obtained. The distance between two points can also be calculated by 
considering elementar analytical geometry. The direct geometrical study of the photo 5 
(Fig.~1) was performed according with the Fig.~2.
%%%%%%%%%%%%%%%%%%%%%%%% FIGURE 2 %%%%%%%%%%%%%%%%%%%%%%%%%%%%%%%%%%%%%%
 \begin{figure}[ht]
 \begin{centering}
 \includegraphics[width=7.5cm]{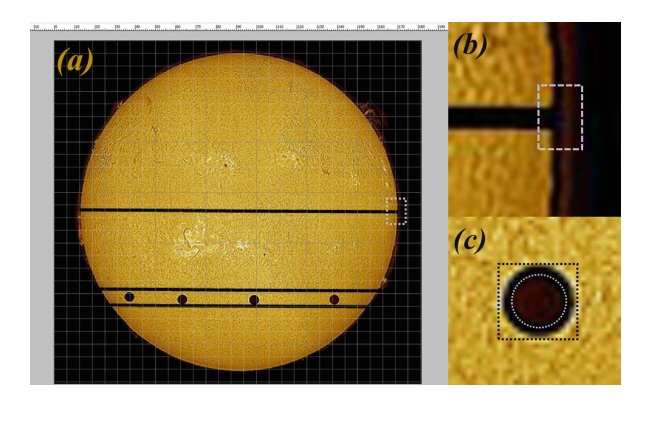}
 \caption[dummy0]{Geometrical treatment of Venus transit of the photo 5 from Fig.~1. 
 In the frame (a), we drew with Photoshop three straight lines, being one the diameter 
 of the Sun and the others enclosing all the spots of Venus  during the transit. 
 In the frame (b), we zoom in the right extreme of the diameter, so that we could 
 securely obtain one point in the illuminated  region and the other one in the dark 
 region. In the frame (c), we zoom in one of the spots, so that it were entirely contained 
 inside a square drawn with the software. Inside the spot, a circle was drawn in order 
 to outline the effective dark region.}
 \end{centering} 
\end{figure}
%%%%%%%%%%%%%%%%%%%%%%%%%%%%%%%%%%%%%%%%%%%%%%%%%%%%%%%%%%%%%%
That figure displays some steps of the procedure adopted to obtain the direct
measurements, using the edition tools of the software. The Fig.~2(a)
displays two parallel straight lines, drawn to enclose all the spots on the Sun. 
The length of the trajectory of Venus spot can be estimated by means of the measurements 
of the lengths of the chords drawn in the figure.   

Above those two lines, a horizontal straight line was traced and the value of the Sun
diameter was obtained. In fact, in agreement with the Fig. 2(b), two values
for the Sun diameter were obtained in that same image. For example, we obtained the 
underestimated value of the Sun diameter considering two points in the extremes of the segment, 
that are entirely contained on the internal illuminated part of the Sun. The
overestimated value of the diameter was obtained choosing the extremes of
the segment on the dark area of the image. It was also measured the value of
the diameter of the spots of Venus in a direct way according to Fig.~2(c),
in which a square encloses the whole spot. A circumference was drawn in
its internal dark region. By means of the square and the circumference, we
obtained two values for the diameter of the spot. All the possible values obtained from 
Fig.~2 are showed in the table 2 and with them we obtained four independent
values for each ratio. 

Now we define two geometrical parameters of the direct model as 
\begin{equation}
\alpha =\frac{D^{\prime }}{D_{S}} \ \ \ \rm{and} \ \ \ \beta =\frac{L^{\prime }}{D_{S}}.  
\label{2}
\end{equation}%
These parameters serve to indicate the quality of the photos and to evaluate the ability of the 
researcher in the procedure of choosing and treating the images. So, we presented in the table 3, %\ref{tab3},  
their average values and their correspondent errors. We also presented in the table 3 %\ref{tab3} 
all the results for the other photos of the mosaic in Fig.~1, that is, %\ref{tab3}, 
we provided the average values of the ratios $\alpha$ and $\beta$, as well their
respective standard deviations $\Delta $ and percentage error $\Delta (\%)$.
This latter value was calculated only to evaluate the quality of the
image. For instance, the parameters $\alpha$ and $\beta$ were numerically obtained by the direct geometrical 
method applied to the best photos and the database were chosen, so that the photos 9 and 12 were discarded 
because they presented high percentage error for the ratio $D^\prime /D_{S}$.

\begin{table}
\caption{Ratios obtained from the photo 5. Each numerical matrix element represents the ratio 
of the diameter or length in the column by the diameter in the row. We considered internal 
values for $D^\prime$ and $L^\prime$, respectively indicated by the first and third columns; 
and external values for both, respectively indicated in the second and fourth columns.}
\begin{center}
\begin{tabular}{|c|c|c|c|c|}
\hline
{\small -} & {\small $D^\prime$} & {\small D}$^{\prime }$ & {\small L}$%
^{\prime }$ & {\small L}$^{\prime }$ \\ \hline
{\small D}$_S$ & {\small 0.03191} & {\small 0.02808} & {\small 0.8596} & 
{\small 0.8239} \\ \hline
{\small D}$_S$ & {\small 0.03197} & {\small 0.02813} & {\small 0.8587} & 
{\small 0.8254} \\ 
\hline
\end{tabular}
\label{tab2}
\end{center}
\end{table}
\begin{table}[bh]
\caption{Average values of the ratios and their respective errors: 
$\Delta$ (standard deviation) and $\Delta$(\%) (percentage error). 
We used the percentage error to study the quality of the images. }
\begin{center}
\begin{tabular}{|c|c|c|c|c|c|c|}
\hline
Photo & $\alpha$ & $\Delta $ & $\Delta (\%)$ & $\beta$ & $\Delta $ & $\Delta (\%)$ \\ \hline
1 & 0.0284 & 0.0028 & 9.7 & 0.8108 & 0.025 & 3.1 \\ \hline
2 & 0.0297 & 0.0024 & 8.0 & 0.8250 & 0.031 & 3.8 \\ \hline
3 & 0.0311 & 0.0009 & 3.0 & 0.8039 & 0.028 & 3.5 \\ \hline
4 & 0.0301 & 0.0017 & 5.7 & 0.8118 & 0.020 & 2.4 \\ \hline
5 & 0.0305 & 0.0020 & 6.7 & 0.8408 & 0.020 & 2.3 \\ \hline
6 & 0.0291 & 0.0038 & 13 & 0.8018 & 0.020 & 2.5 \\ \hline
7 & 0.0326 & 0.0044 & 13 & 0.8220 & 0.012 & 1.5 \\ \hline
8 & 0.0370 & 0.0041 & 11 & 0.8116 & 0.029 & 3.6 \\ \hline
9 & 0.0232 & 0.0038 & 17 & 0.8210 & 0.030 & 3.6 \\ \hline
10 & 0.0239 & 0.0030 & 12 & 0.8187 & 0.023 & 2.8 \\ \hline
11 & 0.0291 & 0.0018 & 6.2 & 0.8637 & 0.030 & 3.5 \\ \hline
12 & 0.0211 & 0.0045 & 21 & 0.8157 & 0.031 & 3.8 \\ \hline
\end{tabular}%
\end{center}
\label{tab3}
\end{table}

Such values are available for being used in the denominated
indirect geometrical method.

\subsection{Indirect Geometrical Method}
The geometrical method is a flat triangulation among the relative positions
of the Earth, the Sun and Venus allowed by transit phenomenon. We here
present three geometrical models, beginning with the simplest one and
increasing the complexity for the others. The results obtained from initial
model have guided us to the proposal of a new complemental model.

A straight line can be imagined instantly connecting the Earth, Venus and its spot on the Sun 
in the beginning of the transit and other in the end of the transit to construct the three triangulations 
of the method according to the Fig. 3. 
%%%%%%%%%%%%%%%%%%%%%%%% FIGURE 3 %%%%%%%%%%%%%%%%%%%%%%%%%%%%%%%%%%%%%%
 \begin{figure}[ht]
 \begin{centering}
 \includegraphics[width=7.5cm]{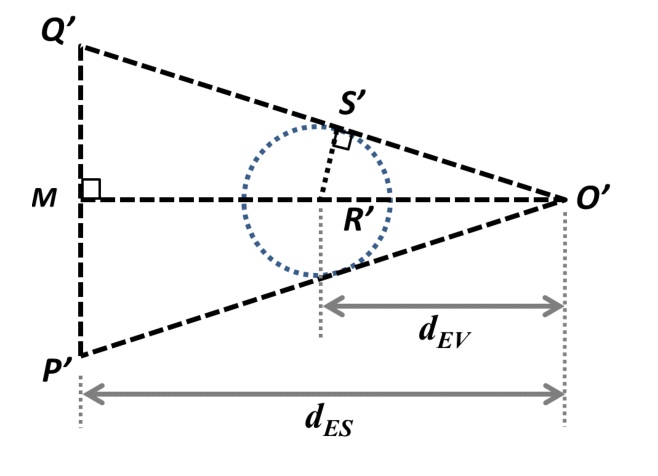}
 \caption[dummy0]{Initial triangulation proposed from Venus transit. In the scheme,  
 we show the basic triangulation, in which is drawn the entire planet and its spot on 
 the Sun. } 
 \end{centering}
\label{fig3}
\end{figure}
%%%%%%%%%%%%%%%%%%%%%%%%%%%%%%%%%%%%%%%%%%%%%%%%%%%%%%%%%%%%%%

In that figure, the triangulation is 
accomplished in the apex of the transit, with Venus, its spot on the Sun and 
one observer in the Earth forming the referred triangle. This is a very special 
instantaneous situation that is always worth. In this triangle, the distance 
$\overline{Q^{^\prime }M}$ \ is the radius of the spot of Venus projected on the Sun,
the distance $\overline{S^{\prime }R^{\prime }}$ is the radius of Venus and  
$O^{\prime }$ is the point in which one finds the observer in the Earth.  
 By using geometry of triangles from that illustration, we found that 
$\Delta Q^{\prime}MO^{\prime } \sim \Delta R^{\prime }S^{\prime }O^{\prime }$,
so that it was obtained the equation
\begin{equation}
\frac{d_{EV}}{D_{V}}=\sqrt{\frac{1}{4}+\frac{1}{\alpha ^{2}}\left( \frac{%
d_{ES}}{D_{S}}\right) ^{2}} .  
\label{4a}
\end{equation}
The numerical value of the parameter $\alpha $\ was obtained by the direct
geometrical method defined by the equation (\ref{2}) and $\frac{d_{EV}}{D_{V}%
}$ can be obtained using the numerical value previously known of the ratio $%
\frac{D_{ES}}{D_{S}}$ given in the table 1.

The Fig. 4 refers to the first model initially proposed, that corresponds to the static case. 
In that model, the translation and rotation movements of the Earth are not considered. 
The entire movement happens as if the Earth and the Sun were
approximately in rest state in relation to the distant stars, during the event~\cite{Feynman1968}. 
 In that figure, the point $P$ is the center of the spot in the
beginning of the transit, when it is just to enter in the solar disk. 
Analogously, $Q$ is the center of the spot in the end of transit. 
Besides, in the interval of time in which the transit happens, we
conjectured Venus movement to be circular uniform one. The triangulation
showed in that figure is mathematically described by
\begin{equation}
d_{EV}=\left[ \frac{2\delta _{V}\frac{d_{ES}}{D_{S}}}{\alpha
+\beta +2\delta _{V}\frac{d_{ES}}{D_{S}}}\right] d_{ES} ,
\label{5a}
\end{equation}
in which 
$\delta _{V}$ is a physical parameter defined by $\delta _{V}=\frac{\pi t}{T_{V}},$
whose numerical value can be obtained from table 1. 
The Eqs.~(\ref{5a}) and (\ref{4a}) can already be enough to obtain a good
value for the Earth-Venus distance and Venus diameter, but it is important to verify the
influence of the Earth revolution movement on the results. So, we proceed
with the translation model.
%%%%%%%%%%%%%%%%%%%%%%%% FIGURE 4 %%%%%%%%%%%%%%%%%%%%%%%%%%%%%%%%%%%%%%
 \begin{figure}[h]
 \begin{centering}
 \includegraphics[width=7.5cm]{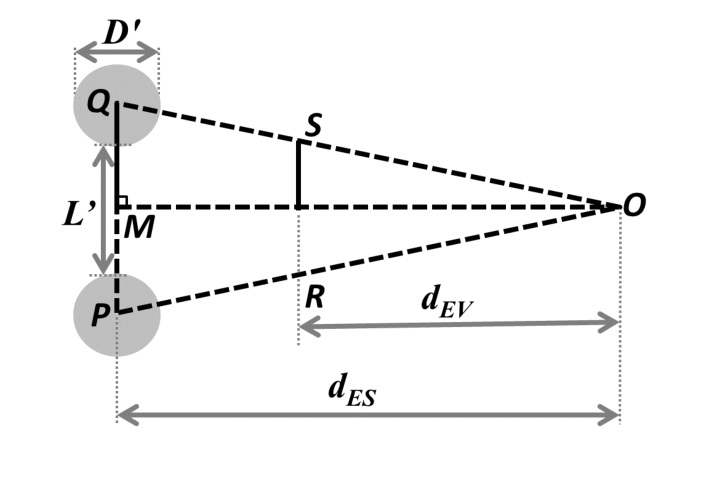} 
 \caption[dummy0]{Triangulation proposed from Venus transit in the case of the first model. In the scheme, 
 we have a representation of the static model, in which it is not considered the Earth revolution movement.} 
 \end{centering}
\label{fig4}
\end{figure}
%%%%%%%%%%%%%%%%%%%%%%%%%%%%%%%%%%%%%%%%%%%%%%%%%%%%%%%%%%%%%%

The Fig. 5 is a triangulation in which the movement of translation of the Earth
around the Sun is added. In that figure, the distance $\overline{TU}$ is the displacement 
of the Earth during the transit. It is important to observe that the
length $d$ is the distance of the Earth up to the point $O$. 
Despite of being an unknown value for us, this does not represent any problem
because it is eliminated in the calculations. 
%%%%%%%%%%%%%%%%%%%%%%%% FIGURE 5 %%%%%%%%%%%%%%%%%%%%%%%%%%%%%%%%%%%%%%
 \begin{figure}[h]
 \begin{centering}
 \includegraphics[width=7.5cm]{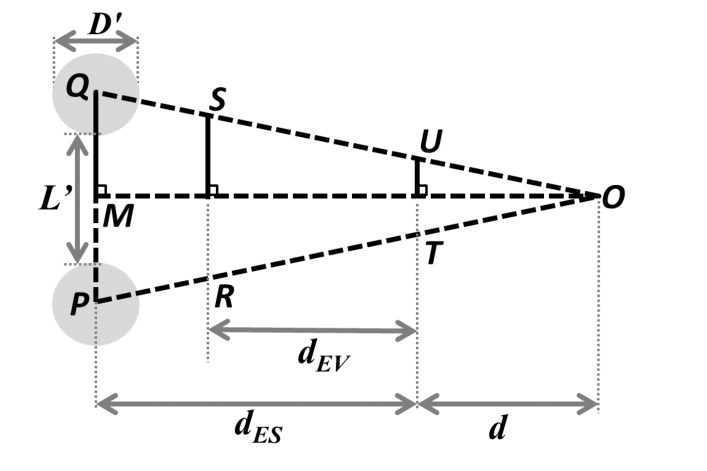} 
 \caption[dummy0]{Triangulation proposed from Venus transit in the case of the second model. In the scheme, 
 we have a representation of the translational model, in which we consider the Earth revolution movement.} 
 \end{centering}
\label{fig5}
\end{figure}
%%%%%%%%%%%%%%%%%%%%%%%%%%%%%%%%%%%%%%%%%%%%%%%%%%%%%%%%%%%%%%

The scheme shown in the Fig. 5 is a complement to the Fig. 4, in which is added the translation 
movement of the Earth. The segment $\overline{RS}$ represents the real trajectory of Venus, while 
the segment $\overline{TU}$ is the trajectory of the Earth during the phenomenon. We propose a flat
model in which the two support straight lines of the segments $\overline{SU}$ and $\overline{RT}$ cross 
each other in the point $O$. That means that the Earth, the Sun and Venus stand together on a plane. 
Inside the triangle $\Delta QMO$, two similar triangles are identified and this lead us to the
equations
\begin{equation}
\frac{\delta _{E}d_{ES}}{d}=\frac{\delta _{V}(d_{ES}-d_{EV})}{(d_{EV}+d)}=
\frac{(\alpha +\beta )D_{S}}{2(d_{ES}+d)} .  
\label{6}
\end{equation}
in which $\delta _{E}=\frac{\pi t}{T_{E}}$
and the numerical value of $T_{E}$ can be obtained from the table 1. The physical parameters $\delta
_{E} $ and $\delta _{V}$ were obtained by surmising uniform circular
movements for the planets. The unknown parameter $d$ can be eliminated in
Eqs. (\ref{6}), then we obtained
\begin{equation}
d_{EV}=\left[ \frac{2\left( \delta _{V}-\delta _{E}\right) \frac{d_{ES}%
}{D_{S}}}{\alpha +\beta +2\left( \delta _{V}-\delta _{E}\right) 
\frac{d_{ES}}{D_{S}}}\right] d_{ES} . 
\label{8}
\end{equation}
The numerical value of Venus diameter was quickly obtained, using the Eq. (\ref{4a}), after the 
results obtained from Eq.~(\ref{8}).
\section{Results and Analysis}
All the values of the distances were calculated using the data from the tables 1 and 3, respectively substituted into Eqs.~(\ref{5a}), 
(\ref{8}) and (\ref{4a}). Our results are showed in the table 4.
\begin{table}[h]
\begin{center}
\caption{Numerical results calculated by means of the indirect geometrical method 
for the static and translational cases. Note that the best results for  
the Earth-Venus distance and Venus diameter were obtained from the translational method. 
The first line of the numerical results refers to the complete transit and the last one corresponds to 
the incomplete one. }
\begin{tabular}{|c|c|c|c|}
\hline
 \multicolumn{2}{|c}{\small{Static}} & \multicolumn{2}{|c|}{\small{Translational}} \\ 
\hline
 ${\mathnormal{\small d}}_{{\small{{EV}}}}${\small (}$ ${\small 10}$^{7}$ {\small km)} & {\small D}$_{{\small V}}${\small (}$ ${\small 10}$^{4}$ {\small km)} & {\small d}$_{EV}${\small (}$ ${\small 10}$^{7}$ {\small 
km)} & {\small D}$_{{\small V}}${\small (}$ ${\small 10}$^{4}$ {\small km)} \\ 
\hline
 {\small{\small{ 7.610}}}$\pm ${\small 0.060} & {\small 2.086}$
\pm ${\small 0.065} & {\small 4.231}$\pm ${\small 0.048} & 
{\small 1.160}$\pm ${\small 0.037} \\ 
\hline
 {\small 7.586}$\pm ${\small 0.097} & {\small 2.11}%
$\pm ${\small 0.30} & {\small 4.187}$\pm ${\small 0.077} & 
{\small 1.16}$\pm ${\small 0.17} \\ \hline
\end{tabular}%
\end{center}
\label{tab6}
\end{table}

By analyzing the data obtained, we can compare the results from the static model with the the corresponding ones from the translational model. We observe that the results calculated from static and translational models were very different  
and the best values of Venus diameter and the Earth-Venus distance were obtained by
means of the translational model, when compared with the values of reference adopted. 
Both values are within of the respective error range provided by the translational model. 
Even so, in order to securely check if it was necessary to correct the results, we
investigated the influence of the Earth intrinsic rotation movement on the
values of table 4. So, we considered the values of the translational model in the 
model with correction due to the Earth rotation.
%%%%%%%%%%%%%%%%%%%%%%%% FIGURE 6 %%%%%%%%%%%%%%%%%%%%%%%%%%%%%%%%%%%%%%
 \begin{figure}[ht] 
 \begin{center}
 \includegraphics[width=6.5cm]{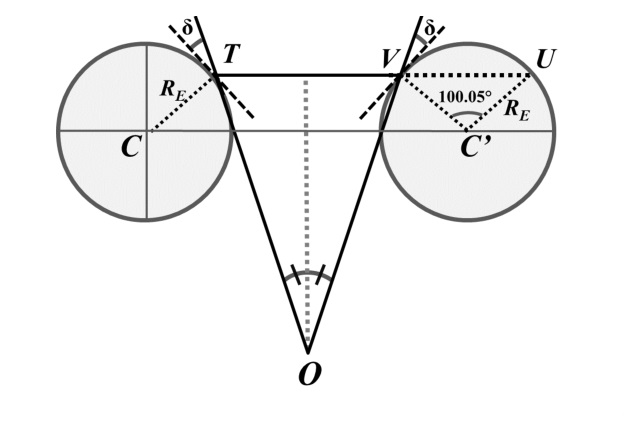}
 \caption[dummy0]{
 Scheme for visualization of the correction in the translational model due to the rotation of the Earth. In 
 the calculation, we first consider a translation of the center of the Earth from C up to C$^\prime$, characterizing the 
 translation $\overline{TU}$. In the end of the translation, one realizes a rotation of $100.05{{}^\circ}$ and the observer in the Earth is dislocated from U to V, characterizing the effective distance $\overline{TV}$. So, the variation in the positions of the observer corresponds to $\overline{UV}=\Delta l$. In both circles, the parameter $\delta$ is the angle of observation in relation to the horizon.}
\end{center}
\end{figure}

We decomposed the movement of the Earth in two sequential movements in order to obtain the correction. The first one was the translation of the Earth and afterwards the rotation motion.
The Fig.~6 shows the triangulation of the transit when it is considered that additional movement. 
The first circle to the left represents the beginning of the transit and the right one corresponds to its end.  
The radius of the Earth $R_E$ appears in a natural way as a parameter of correction, with value given in table 1. 
In that figure, we can analyze a scheme of a particular symmetrical configuration for the correction in the translational model due to the rotation of the Earth. In the circles representing the Earth, the point $T$ indicates the position of the observer in the beginning of Venus transit and $V$ is the position of the observer in the end of transit, taking into account both movements. We recognize the translation distance $\overline{TU}$, earlier used in the translational model. The angle of rotation of the Earth that corresponds to the time of transit is $100.05{{}^\circ}$ and $\overline{TV}$ is a corrected distance due to the new position of the observer after the rotation of the Earth, so that the difference between those positions of the observer  corresponds to $\overline{UV}=\Delta l$. We also see in both circles  the angle $\delta$ of observation of the phenomenon in relation to the horizon. 
The correction $\overline{UV}$\ can be obtained from the equation $\overline{TV}=2\delta_{E}d_{ES}-\Delta l.$ 

In such a symmetrical configuration, one can in a straightforward way obtain the correction $\Delta l = 2R_E\sin 50.025^\circ$, so that 
\begin{equation}
d_{EV}=\left\{ \frac{\left[ 2\left( \delta _{V}-\delta _{E}\right) +%
\frac{\Delta l}{d_{ES}}\right] \frac{d_{ES}}{D_{S}}}{\alpha +\beta +%
\left[ 2\left( \delta _{V}-\delta _{E}\right) +\frac{\Delta l}{d_{ES}}\right]
\frac{d_{ES}}{D_{S}}}\right\} d_{ES} .  
\label{20}
\end{equation}%
So, we calculated the corrected values by considering the symmetrical configuration and our final values 
are shown in the table 5. 
\begin{table}[h]
\caption{Final results for the Earth-Venus distance and Venus diameter by means of the translational 
model with correction. }
\begin{center}
\begin{tabular}{|c|c|}
\hline
$d_{EV}(km)$ & $ D_{V} (km) $ \\
\hline
$4.296 \ \pm $ $0.048$ & $1.178 \pm 0.038$  \\ 
\hline
\end{tabular}
\end{center}
\label{table6}
\end{table}

The corrections had a small influence on the values of the table 4, mainly on the average values, but these 
corrected values are closer to the values of reference adopted.  
\section{Concluding remarks}
This work achieved the objective of obtaining the numerical values of Venus diameter and the Earth-Venus distance by 
means of models based on image treatment and triangulations allowed by Venus transit.  
The proposal highlights the efforts from several researchers and observers that 
obtained the photos used in the geometrical treatment of the images and presents as advantage to consider simple flat 
geometrical models in the description. In fact, we completely reached our aims by means of the translational model with 
corrections, within a relatively good precision. In the calculations, we adopted as input data two geometrical parameters obtained from a set of photos with the composition of the transit. In the case of the rotational correction, the Earth radius 
appeared in a natural way as a correction parameter. The final results took into account the systematic error due to 
the equipments, such as the type of cameras with its respective coupled filters, and errors in the composition of
images. The geometrical methods here proposed depend on the researcher and his choices, which can present small 
variations from one to the other (as evaluations of spots diameter or positions of extreme points in diameters), 
but the method is itself robust, with all the values laid within the error bar. 
It is important to mention that the average values of the Earth-Venus distance and their respective percentage errors 
were more precise, but in the case of Venus diameter the error propagation significantly increased the percentage error 
in the incomplete transit. Despite of that problem, we did not discard them because the average values were very close to 
the measurements of reference, so that one could identify the differences between the results of the 
complete and incomplete transit. 
We believe that the new technique outlined here can be applied to other planets and other Moons, as in the case of the transits 
of Jupiter Moons. In the case of an exoplanet, we could determine the distance between it and its respective star if we knew the  
apparent diameter of the star, the distance between the star and the Earth and the composition of the transit, so that the method 
here described should work well. However, to analyze distances or orbital parameters of those orbs we need additional investigations that are 
now in progress. 

\section*{Acknowledgments}

We thank to all the researchers responsible for the photos~\cite{Mosaic2012} that allowed us to construct the mosaic used in this work. Victo S. Filho thanks to S\~ao Paulo Research Foundation (FAPESP) for partial financial support.


\begin{thebibliography}{99}

\bibitem{Dick2004} S. J. Dick, The Transit of Venus, Scientific American 290, 98-105 (2004)  

\bibitem{Pasachoff2012} J. M. Pasachoff, Transit of Venus: Last chance to see, Nature 485, 303-304 (2012) 

\bibitem{Hind1871} J. R. Hind, The Transits of Venus in 2004 and 2012, Nature 3, 513-514 (1871) 

\bibitem{Forbes1879} G. Forbes, The Coming Transit of Venus, Nature 9, 447-449 (1879)

\bibitem{Argyll+1882} R. S. B. Argyll {\it et al.} The Transit of Venus, Nature 27, 154-159 (1882) 

\bibitem{Meadows1974} A. J. Meadows, The transit of Venus in 1874, Nature 250, 749-752 (1974)  

\bibitem{Goody1967} R. Goody, The scale height of the venus haze layer, Planetary and Space Science 15, 1817-1819 (1967)

\bibitem{Chassefierea2012} E. Chassefi{\`{e}}rea, R. Wielerc, B. Martyd and F. Leblance F., The evolution of Venus: Present state of knowledge and future exploration, Planetary and Space Science 63–64, 15–23 (2012)

\bibitem{Heilmann} A. Heilmann, L. D. D. Ferreira, C. A. Dartora, The Effects of an Induced Electric Dipole Moment due to Earth’s Electric Field on the Artificial Satellites Orbit, Brazilian Journal of Physics 42, Issue 1-2, 55-58 (2012)

\bibitem{Suspes} Framsol L\'opez-Suspes, Guillermo A. Gonz\'alez, Equatorial Circular Orbits of Neutral Test Particles in Weyl Spacetimes, Brazilian Journal of Physics 44, Issue 4, 385-397 (2014)

\bibitem{Bolos} V. J. Bol\'os,  Kinematic relative velocity with respect to stationary observers in Schwarzschild spacetime,  J. Geom. Phys. 66, 18 (2013)

\bibitem{Pugliese} D. Pugliese, H. Quevedo, R. Ruffini, Equatorial circular motion in Kerr spacetime, Phys. Rev. D 84, 
 044030 (2011) 

\bibitem{Mosaic2012} The photos used in the construction of the mosaic
in Fig.~1 were: Photo 1: Joetsu City (Niigata, Japain), 06 June 2012; Photo 2:
Manila (Philippines). Credit: James Kevin, 06 June 2012; Photo 3: Quezon City
(Philippines). Credit: Jett Aguilar, 06 June 2012; Photo 4: Mauna Loa (HI, USA).
Credit: National Solar Observatory; Photo 5: Mauna Kea (HI, USA), 05 June
2012, H$\alpha $ Composite Image; Photo 6: Landers (CA, USA), 05 June 2012;
Photo 7: Credit: \copyright\ Indranil Sinharoy; Photo 8: Mount Wilson
Observatory. Credit: Jie Gu; Photo 9: Port Angeles (Wash, USA). Credit: Rick
Klawitter, 05 June 2012; Photo 10: Scottsdale (Arizona, USA). Credit: 
\copyright\ 2012 Brian Leckett; Photo 11: Abu Dhabi. Credit: Nik Syahron, 06
June 2012; Photo 12: Sta Rosa (Laguna, Philippines). Credit: Doun
Dounel, 06 June 2012.

\bibitem{Stone+1882} E. J. Stone, S. P. Langley and J. Birmingham, Transit of Venus, Nature 27, 177-180 (1882) 

\bibitem{Hughes2003} D. W. Hughes, Planetary Spin. Planetary and Space Science 51, 517-523 (2003)  

\bibitem{Halliday2012} D. Halliday, F\'isica, Vol. 1 (LTC, Rio de Janeiro, 2012)

\bibitem{pdg2010} Particle Data Group, J. Phys. G 37, 102 (2010)  

\bibitem{Brown1998} T. M. Brown and J. Christensen-Dalsgaard, Accurate determination of the solar photospheric radius, Astrophys. J. 500, L195 (1998) 

\bibitem{Espenak2014} F. Espenak, Eclipses and the Moon's Orbit, NASA's GSFC (2012) http://eclipse.gsfc.nasa.gov/SEhelp/moonorbit.html. Access: March, 2015

\bibitem{Feynman1968} R. P. Feynman, R. B. Leighton and M. Sands, The Feynman Lectures on Physics,  Vol. I (Addison-Weslley, Massachussetts, 1968)

\end{thebibliography}
\end{document}